# Performance analysis of InAlN/GaN HEMT and optimization for high frequency applications

Jagori Raychaudhuri[a,b], Jayjit Mukherjee[b], Amit Malik[b], Sudhir Kumar[b] D.S. Rawal[b], Meena Mishra[b,*], Santanu Ghosh[a]

[a]Department of Physics, Indian Institute of Technology, Delhi, Hauz Khas, Delhi-110016, India

[b]Solid State Physics Laboratory, DRDO, Timarpur, Delhi-110054, India

*meena.mishra23@yahoo.com

**Abstract:**

An InAlN/GaN HEMT device was studied using extensive temperature dependent DC IV measurements and CV measurements. Barrier traps in the InAlN layer were characterized using transient analysis. Forward gate current was modelled using analytical equations. RF performance of the device was also studied and device parameters were extracted following small signal equivalent circuit model. Extensive simulations in Silvaco TCAD were also carried out by varying stem height, gate length and incorporating back barrier to optimize the suitability of this device in Ku-band by reducing the detrimental Short Channel Effects (SCEs). In this paper a novel structure i.e., a short length T gate with recess, on thin GaN buffer to achieve high cut-off frequency ($f_T$) and high maximum oscillating frequency ($f_{max}$) apt for Ku-band applications is also proposed.

1. **Introduction:**

In the era of uprising demand of high power, high efficiency and high frequency [1,2,3] applications GaN is the most suitable candidate to deliver outstanding results since it is blessed with some superior qualities like wide bandgap, high saturation velocity etc [4,5]. AlGaN/GaN HEMTs are extremely popular as their RF performance is very good [6,7]and they are well suited for high power applications [8]. But we cannot neglect the reliability issues that they face due to the strain from the lattice mismatch [9 10]. With the increasing demand of high frequency applications several modification techniques were applied on AlGaN/GaN HEMTs. Scaling down the gate length is one of the solutions to increase the frequency of operation as there is an inverse relationship between cut off frequency and gate length [11,12]. But with drastic reduction of gate length, the short channel effects (SCE) arise and degrade the performance of the device [13,14,15]. Maintaining high aspect ratio ($L_g/t_{bar}$) can resolve the issue of SCEs [16,17,18] but thinning down the barrier has shown surface depletion problems in AlGaN/GaN HEMT [19]. In the context of high frequency applications some new thin barrier heterostructures has come to play [20,21,22]. InAlN/GaN is one of the most promising contenders among them. Strain free lattice matched layer structure with 17% In, makes it more reliable [23,10,23] by reducing defects. High thermal stability is also reported in some works [24]. Since these devices are thin barriered, in smaller gate length devices suppression of SCEs has been achieved up to a great extent. InAlN devices are showing good results in high frequency [25,26,27] high power and high temperature [24] applications.

But SCEs still persist when gate lengths are scaled down in the ultrashort regime. Thinning down top barrier is not a practical solution as it increases leakage and reduces 2DEG [26]. Introduction of Back barrier came to rescue [26,28], both InGaN and AlGaN back barriers helped to enhance RF performance of the device with sub-micron gate lengths [26,28,29]. With gate length, gate height of a T-gate also plays a role to determine the device performance. An optimised stem height can manifest good results reducing gate capacitances and gate resistances [30,31]. Despite having such remarkable material characteristics GaN is heavily vulnerable to traps [32, 33,34,35]. Barrier traps in InAlN/GaN is also reported [36,37] in several literatures that limit device performances.

In this paper we have studied an inhouse grown and fabricated HEMT with lattice matched InAlN barrier. We have examined the DC and RF characteristics of the device. The barrier traps were also characterised by Current- Deep Level Transient Spectroscopy (C-DLTS) technique. We have optimised the gate length and gate height to obtain better performance of the device in higher frequency regime by conducting simulation study on the same structure. The SCEs were also observed and we have studied how to mitigate them using an AlGaN backbarrier. Further from our simulations we have proposed submicron gate length with recess, on a thin GaN buffer with InAlN barrier, which shows promising results in high frequency applications.

## 2. Sample Details:

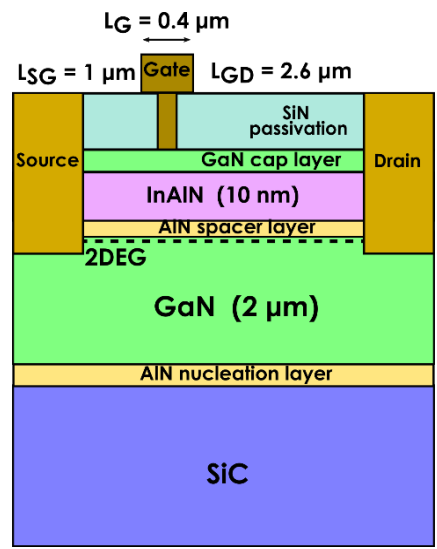

Fig. 1 Layer structure of the device.

The layer structure of the inhouse grown device is shown in Fig. 1. The epistructure was grown by MOCVD technique. 10 nm thick InAlN layer with 17 % In was grown on 2μm GaN layer. The GaN buffer is grown on SiC substrate. 1 nm of AlN spacer layer was used. The channel in the heterostructure has a sheet charge density of $1.85 \times 10^{13}$ cm$^{-2}$ and electron mobility of 1780 cm$^2$/V.s. For device fabrication the source, drain ohmic contacts were realized using Ti/Al/Ni/Au metal stack followed by rapid thermal annealing. The MESA isolation was carried out using inductively coupled plasma reactive ion etching (ICPRIE) with Cl$_2$/BCl$_3$ chemistry. Schottky gate contacts were realised using Ni/Au metal scheme. The gate length of the fabricated device was 400 nm with 100 nm stem height and gate width of 100 μm. The device was passivated with a 100 nm layer of plasma enhanced chemical vapour deposition (PECVD) grown SiN.

### 3. Experimental Details:

DC-IV and C-V characterizations were done using Keithley SCS 4200A semiconductor parametric analyzer. Output and transfer characteristics were measured. Temperature dependent DC measurements were conducted up to 80°C. C-V measurements were done. The barrier traps were characterized and forward gate current was modelled. S-parameter measurements were also carried out at different bias conditions to understand RF performance of the device. From the measured S-parameter data $f_T$ and $f_{max}$ were calculated and the device parameters were extracted following small signal equivalent circuit model. Simulation study was conducted in Silvaco TCAD to improve the device performance in higher frequency range.

### 4. Results and discussion:

**4.1 DC -IV Characterization of the Device:**

From the DC-IV transfer characteristics, the maximum transconductance, $g_{m,max}$ is obtained as 430 mS/mm and from the DC-IV output characteristics, it is found that the device exhibits high saturation current (1 A/mm) (Fig. 2(a) and 2(b).) which reflects high 2DEG density yielded from matched epistructure.

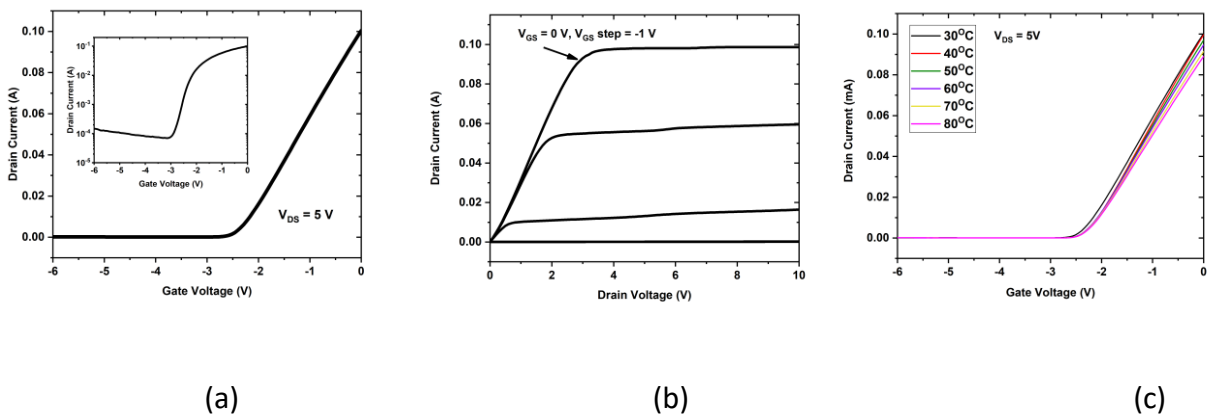

(a) (b) (c)

Fig. 2 DC-IV characteristics of InAlN/GaN device (a) Transfer characteristics measured at $V_{DS}$ = 5 V, inset shows the plot in logarithmic scale (b) Output characteristics (c) Transfer characteristics for a temperature range of 30°C – 80°C.

Temperature dependent DC-IV measurements were done over the range of 30°C to 80°C. From this measurement, negligible shift (0.12 V) in the threshold voltage ($V_T$) indicates towards less accumulation of barrier traps (Fig. 2(c)).

**4.2 Trap Characterization using C-DLTS:**

To observe trapping and to calculate the activation energy of the traps in barrier, we have used a filling pulse of 100 sec with $V_{GS}$ = -2.4 V and $V_D$ = 3 V ensuring that the depletion region extends only into the barrier. For the detrapping phase, $V_{GS}$ = 0 V and $V_{DS}$ = 3 V is applied for 1000 sec. The entire experiment is repeated up to 80°C in step of 10°C. The plot of the derivative of the drain current with respect to $\log_{10}(t)$ vs $\log_{10}(t)$ shows the dominant time constants responsible for the detrapping transients. With increase in temperature a leftward shift in the dominant time constants is observed (Fig. 3(a)). The Arrhenius formulation is utilized to extract the trap energy and it was calculated to be 0.29 eV from the conduction band (Fig. 3(b)).

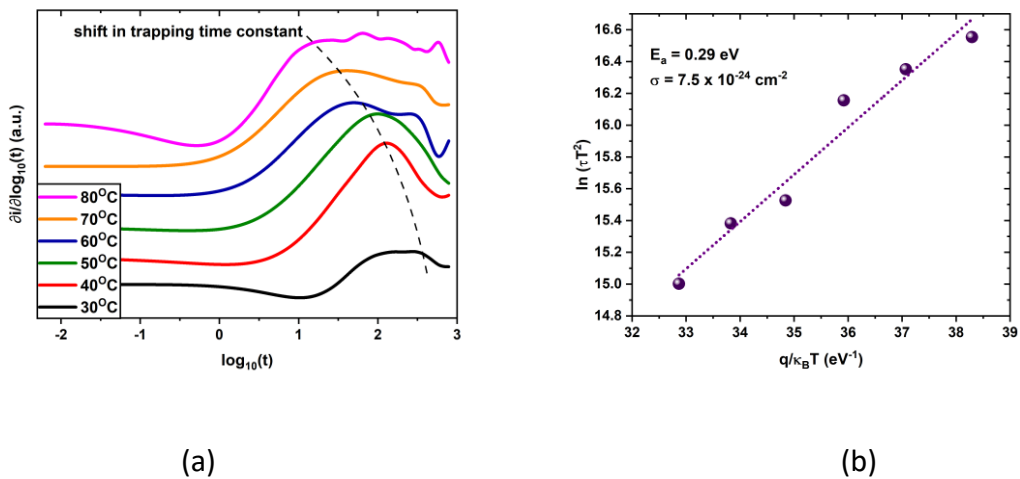

(a)                                     (b)

Fig. 3 (a) C-DLTS spectra of the detrapping transients for different temperatures (the dotted line denotes the peak shifting with increasing temperature) (b) Arrhenius plot to calculate activation energy of the traps extracted from C-DLTS spectra

**4.3 Modelling of Forward Gate Current:**

The forward gate current was showing the signatures of multiple processes (Fig. 4). In previous works the gate current was modelled following different mechanisms [38,39,40]. Here we have considered Thermionic Emission (T.E.), Trap assisted Tunneling (T.U.) and leakage (R.L.) components to model the forward gate current.

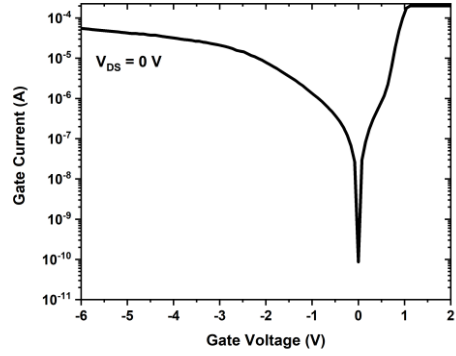

Fig. 4 Gate current measured at $V_{DS}$=0V

To realize the forward gate current, we have used the following equations regarding the 3 components [42],

$$I_{TE} = I_{TE}(0)\left[\exp\left(\frac{q(V-IR_S)}{\eta\kappa T}\right) - 1\right] \quad (1)$$

$$I_{TU} = I_{TU}(0)\left[\exp\left(\frac{q(V-IR_S)}{E_T}\right) - 1\right] \quad (2)$$

$$I_{RL} = \frac{V-V_0}{R_L} \quad (3)$$

$$I_{TE}(0) = AA^*\frac{T^2\phi_B}{\eta\kappa T} \quad (4)$$

$$I_{TU}(0) = qv_D N_{dis}\exp(-\frac{\phi_B}{\kappa T}) \quad (5)$$

Where $I_{TE}$ (0), $I_{TU}$ (0) are components at zero bias, $R_L$ is the leakage resistance, $R_s$ is the series resistance, T is the absolute temperature, $E_T$ is the tunneling energy, $\phi_B$ is the Schottky barrier height and $\eta$ is the ideality factor A is the Schottky area of the device and A* is the Richardson constant, $v_D$ is Debye frequency and $N_{dis}$ is dislocation density.

To realize the forward current first, schottky barrier height $\phi_B$ and ideality factor $\eta$ were calculated over the temperature range (Fig. 5). From Fig. 5(b) the ideality factor shows a higher value ($\eta \approx 3 - 3.28$) over the experimental temperature regime. The high value of $\eta$ indicates the existence of defect assisted tunneling along with the thermionic emission processes in the gate current [41, 42]. Hence, tunneling was considered with thermionic emission and also combining these two mechanisms with leakage, the forward gate current of our device was successfully modelled (Fig. 6).

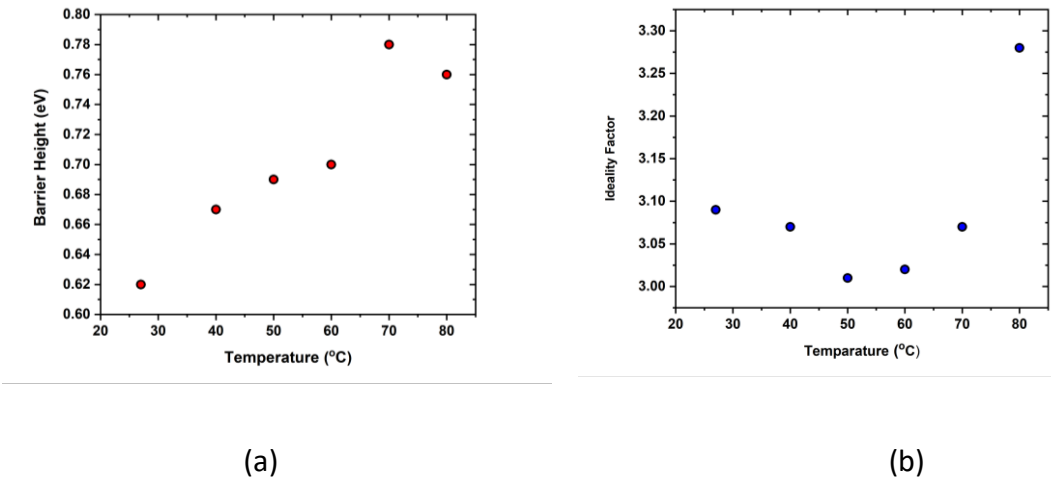

(a) (b)

Fig. 5 Variation of (a) Barrier Height and (b) ideality factor with temperature

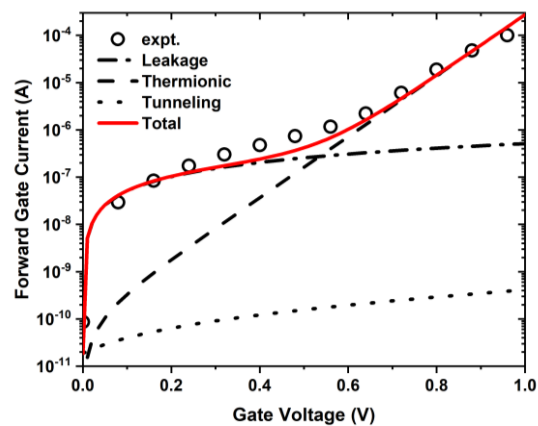

Fig. 6 Modelled Forward gate current using thermionic emission, tunneling and leakage

**4.4 C-V Characterization:**

From multi frequency C- V measurement interface trap state ($N_{it}$) and trap state density ($D_{it}$) can be calculated considering the following equations [43]

$$N_{it} = C_b \Delta V / q \tag{1}$$

$$D_{it} = N_{it} / \Delta E_t \tag{2}$$

$$E_t = \kappa T \ln\left(\frac{f_1}{f_2}\right) \tag{3}$$

Where $C_b$ is the barrier capacitance, $E_t$ is the energy difference between the traps, $f_1$, $f_2$ are two different frequencies which are 1 MHz and 500 kHz for our experiment and $\Delta V$ is the voltage difference for two different frequencies corresponding to the barrier capacitance.

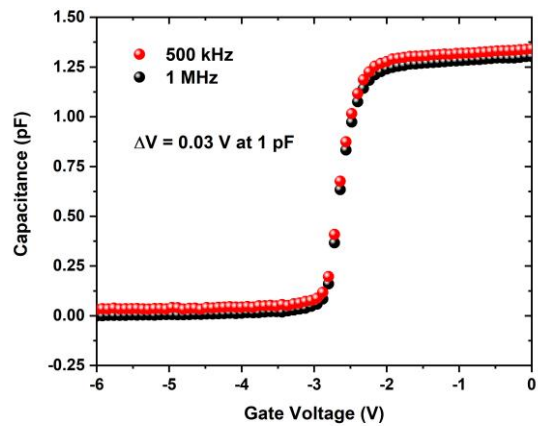

Fig. 7 Experimental Capacitance- Voltage Characteristics at 500 kHz and 1 MHz

From experiment following these equations (Fig 7). $N_{it}$ =1.5 X $10^{11}$ /cm² and $D_{it}$ = 8.3 x$10^{12}$ /eV cm² are obtained. These results denote a good quality interface with less population of interface traps.

**4.5 RF Characterization:**

Cut-off frequency ($f_T$) and maximum oscillation frequency ($f_{max}$) from unity current gain ($h_{21}$) and Maximum gain (MAG/MSG) using S-parameter measurements at the bias point ($V_{GS}$ = -1.5 V, $V_{DS}$ = 10 V) were calculated (Fig. 8).

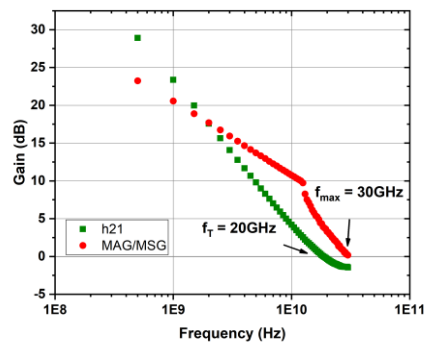

Fig. 8 $f_T$ and $f_{max}$ as calculated from the unity current gain ($h_{21}$) and maximum gain (MAG/MSG) of the device

$f_T$ = 20 GHz and $f_{max}$ = 30 GHz are extracted from measured S-parameters for the device. It is much less than what we expect from InAlN/GaN devices. In our device we have used a T-gate with stem height of 100 nm. A 100 nm stem height on a 10 nm barrier can induce increment in gate capacitances which in turn degrades the RF performance of the device.

To understand this, the device parameters were extracted from small signal equivalent circuit using measured S-parameters [44,45,46]. First the extrinsic parameters were calculated, then the intrinsic parameters were obtained. The list of parameters is shown in Table 1.

| | Parameters | Value |
|---|---|---|
| **Extrinsic** | $R_g$ (Ω) | 5.02 |
| | $R_d$ (Ω) | 9.18 |
| | $R_s$ (Ω) | 11.1 |
| | $L_d$ (pH) | 76 |
| | $L_g$ (pH) | 28.4 |
| | $L_s$ (pH) | 1 |
| | $C_{pg}$ (fF) | 100.75 |
| | $C_{pd}$ (fF) | 44.8 |
| **Intrinsic** | $g_m$ (mS) | 62.5 |
| | $g_d$ (mS) | 7.46 |
| | $R_i$ (Ω) | 1.4 |
| | $C_{gd}$ (fF) | 52 |
| | $C_{gs}$ (fF) | 285 |
| | $C_{ds}$ (fF) | 16.5 |
| | $\tau$ (ps) | 1.75 |

Table 1. Extracted device parameters using small signal equivalent circuit model for the device.

The table also indicates the increased value of gate capacitances which compromises the operating frequency of the device.

### 4.6 Simulation Study:

To understand the effect of structural parameters of the device on high frequency performances we have done a simulation study in Silvaco TCAD. We have varied device parameters, incorporated special structures to optimize the device performance. The experimental DC-IV (transfer, leakage and output) characteristics of the device was calibrated with the simulated results. The layer structure and device dimensions were kept same as our physical device as shown in Fig. 1. The physical models that were used are Fermi statistics, Shockley-Read-Hall (SRH) recombination, UST and GaNSat, Fldmob for mobility and Pipinys for the gate leakage. Fig. 9. shows the calibrated characteristics.

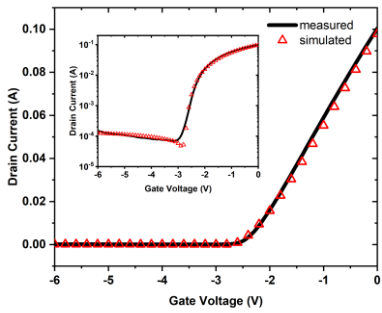 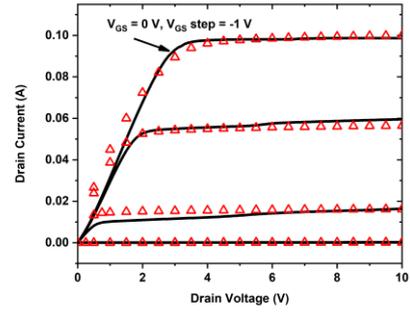

(a) (b)

Fig. 9 Calibration of experimental device characteristics with simulated results for the InAlN/GaN HEMT (a) Transfer characteristics (b) output characteristics

To understand how stem height impacts the optimum RF performance of the device, a simulation study was carried out by varying the stem height and observed the operating frequencies keeping the gate length 400 nm (Fig. 10).

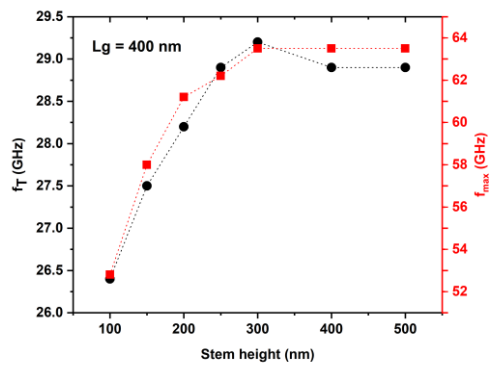

Fig. 10 Variation of $f_T$ and $f_{max}$ with gate stem height in InAlN/GaN HEMT

Starting from 100 nm, it is observed that $f_T$ and $f_{max}$ were improving up to 300 nm stem height, beyond that the results are deteriorating due to the effect of gate resistances and gate capacitances.

A popular way to increase the cut off frequency is scaling down the gate length. Therefore, we have scaled down the gate length up to 30 nm for different stem heights (Fig. 11). From Fig

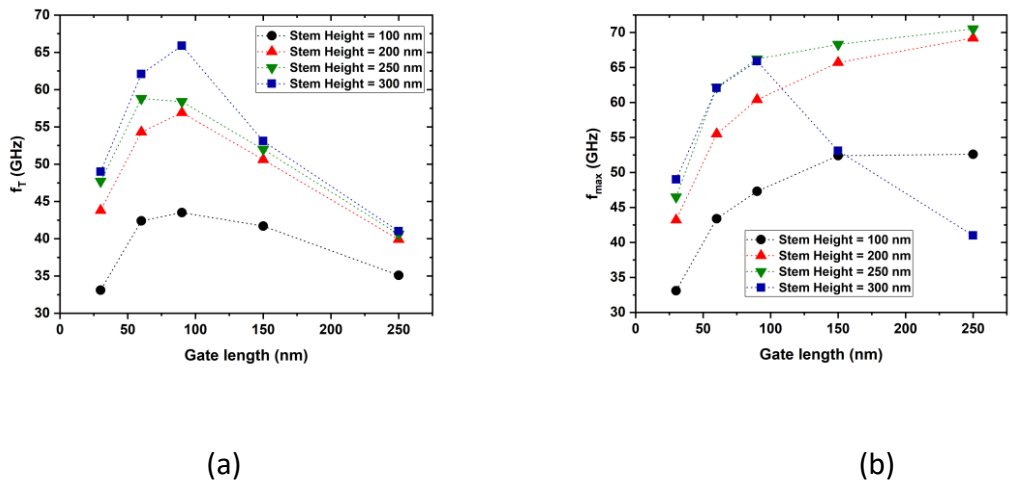

(a)                                                                 (b)

Fig. 11 (a) Variation of $f_T$ and (b) $f_{max}$ with gate length in InAlN/GaN HEMT

11(a) it is observed that for all the stem heights, the $f_T$ increases upto 90 nm gate length after which it deteriorates on further scaling. We are getting the highest $f_T$ with a combination of 90nm gate length and 300 nm stem height. However, from figure 11(b) it is observed that the $f_{max}$ is again highest upto 90 nm gate length. But when it comes to stem height 250 nm yields better $f_{max}$ with 90 nm gatelength.  Thus from the point of view of both $f_T$ and $f_{max}$ , T- gate with 250 nm stem height and 90 nm gate length was optimized. Fig 12. shows the frequency performance of the T- gate with 250 nm stem height and 90 nm gate length.

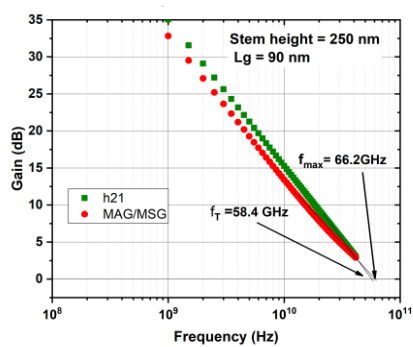

Fig. 12    $f_T$ and $f_{max}$ with T gate having 90nm length and 250nm stem height in InAlN/GaN HEMT

Since drastic cutting down of gate length invokes Short Channel Effects, scaling down to 60nm and 30 nm gate lengths doesn't yield better frequency performance.

To examine this, Drain Induced Barrier Lowering (DIBL) was calculated for different gate lengths keeping stem height 250 nm. From Fig 13 DIBL is quite evident in the cases of 60 nm and 30 nm gate lengths.

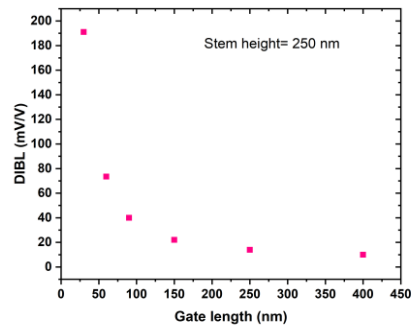

Fig. 13 Variation of DIBL with gate length in InAlN/GaN HEMT

To suppress SCEs an 850nm $Al_{.25}Ga_{.75}N$ back barrier was incorporated keeping 10 nm GaN channel on the same calibrated structure (Fig. 14(a)). Frequency performance improves even in 60nm and 30 nm gate length devices (Fig. 14(b)) as the back barrier suppresses the DIBL (Fig. 14(c)). Thus, scaling down can be successfully achieved with 30 nm gate length with AlGaN back barrier. Fig. 15 depicts good frequency performance with the above optimized structure.

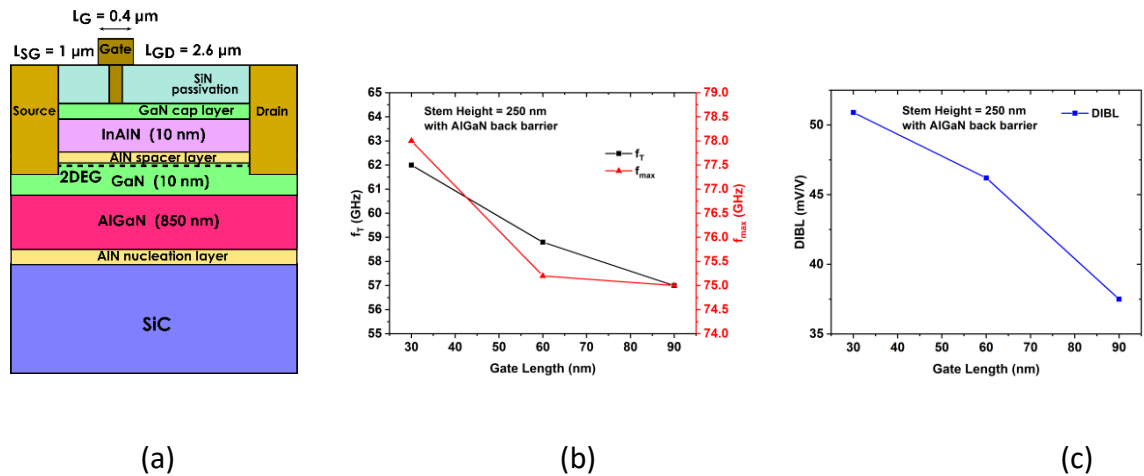

          (a)                         (b)                         (c)

Fig14. (a) Layer structure with backbarrier (b) Variation of $f_T$, $f_{max}$ and (c) Variation of DIBL with submicron gate length in InAlN/GaN HEMT with AlGaN back barrier.

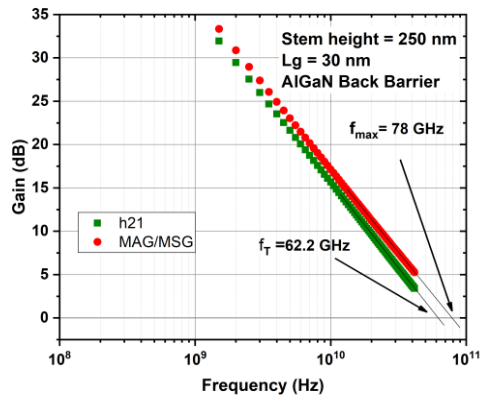

Fig.15 $f_T$ and $f_{max}$ of T gate having 30nm length and 250nm stem height in InAlN/GaN HEMT with AlGaN backbarrier.

Buffer free or thin buffer AlGaN/GaN structures are hot topic for discussion these days as they are showing promising results in terms of device performance [47,48]. It would be a nice idea to examine thin barrier structure on thin buffer structure. It would increase the throughput and incorporate the thin buffer related benefits into the device. Keeping this in mind, the same barrier layer on a thin GaN buffer of 200 nm thickness was simulated with the stem height at 250 nm for sub-100 nm gate length devices. The performance of thin buffer structure is almost similar to that of the conventional buffer. It shows appreciable frequency response with 90 nm get length but unlike the backbarrier structure 60nm and 30 nm show the degradation of RF performance due to the severity of SCEs. Comparative study of $f_T$, $f_{max}$ and DIBL for different gate lengths (sub-100 nm) on the thin GaN buffer structure is depicted in Fig. 16.

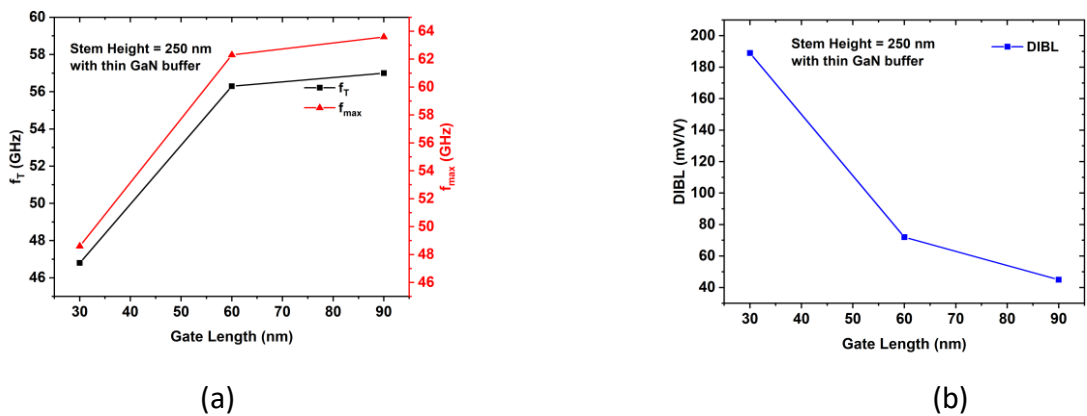

(a)                                      (b)

Fig. 16 (a) Variation of $f_T$, $f_{max}$ and (b) variation of DIBL with submicron gate length in InAlN/GaN HEMT on 200nm thin GaN buffer

Another way to combat SCEs is to use the recessed gate structure as it improves the aspect ratio (Lg/t$_{bar}$) of the device [49]. A recessed gate on the thin GaN buffer layer structure is incorporated and a comparative simulation study is carried out for different sub-100 nm gate lengths. We have optimized the recess depth of 3 nm into the InAlN barrier to keep the 2DEG intact and to prevent further leakage (Fig. 17(a)).

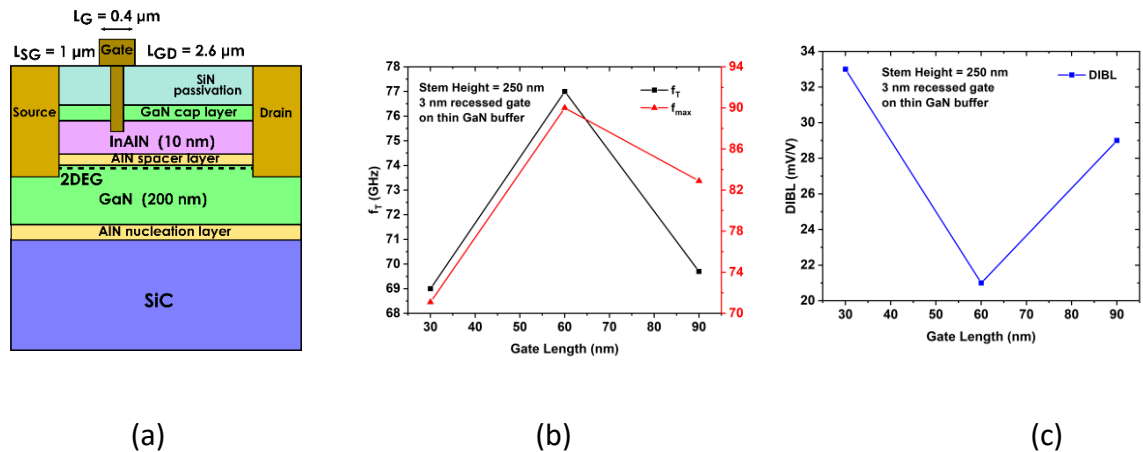

(a)                  (b)                  (c)

Fig. 17 (a) layer structure grown on thin GaN with recessed gate (b) Variation of f$_T$, f$_{max}$ and (c) Variation of DIBL with submicron gate length with 3 nm recess in InAlN/GaN HEMT on 200nm thin GaN buffer

From Fig.17(b) the improvement in f$_T$ and f$_{max}$ is evident. From Fig.17(c) it is observed that the DIBL is reduced to a great extent in the case of 60 nm gate length. Although the DIBL increased in 30 nm gate length device but it didn't affect the frequency performance much. A recessed gate structure having 60 nm gate length and 250 nm stem height on InAlN barrier and thin GaN buffer heterostructure has been optimized. This device achieves a suitability in high frequency operation with f$_T$ = 77 GHz and f$_{max}$ = 90 GHz as shown in Fig. 18.

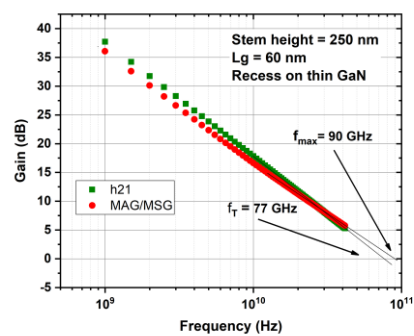

Fig.18 f$_T$ and f$_{max}$ of T-gate having 60nm length and 250nm stem height with 3 nm recess in InAlN/GaN HEMT with 200nm thin GaN buffer.

## 5. Conclusion:

It is found that the InAlN/GaN HEMT device is yielding appreciable amount of drain current and transconductance due to its profound 2DEG density. From C-V analysis and transient characterizations insight about interface and barrier traps was determined. It is introspected how different mechanisms impact on forward gate current using thin barrier InAlN structure. The simulation study reveals the effect of stem height and gate length on $f_T$ and $f_{max}$. T-gate with stem height of 250 nm with gate length of 90 nm shows high cut- off frequency of 58.4 GHz. With the incorporation of a 850 nm AlGaN back barrier, the short channel effects were suppressed and a higher cut -off frequency (62.2 GHz) using gate length of only 30 nm was obtained. The thin barrier of InAlN grown on thin GaN buffer has emerged as a potential contender for high frequency applications from the simulation study. A 3nm recessed T gate with length of 60 nm simulated on thin GaN buffer shows a promising result ($f_T$ = 77 GHz, $f_{max}$ = 90 GHz) in Ku-regime. This gives a scope to fabricate and validate this structure in future.

**Acknowledgment**

The authors would like to thank GaN Team and Director SSPL, DRDO, Delhi for the valuable support and guidance to carry out the experimental work.

**Data Availability Statement**

The data that support the findings of this study are available upon reasonable request from the authors.

**Conflict of interest**

The authors declare that they have no known competing interests.